%% file: BPtraffic.tex
\documentclass[11pt,a4paper,twoside]{paper}

\usepackage{ae,aeguill,textcomp}
\usepackage[latin1]{inputenc}

\newif\ifRR\RRtrue
\ifRR
  \usepackage{RR}
\else
  \usepackage{graphicx}
\fi

\usepackage{amsmath,amssymb,amsthm}

\newcommand{\G}{\mathcal{G}}
\newcommand{\V}{\mathcal{V}}
\newcommand{\E}{\mathcal{E}}

\makeatletter

\@addtoreset{equation}{section}

\@addtoreset{figure}{section}

\@addtoreset{table}{section}
\makeatother

\makeatletter
\@ifundefined{theorembodyfont}{}{
 \newtheoremstyle{paper}{\topsep}{\topsep}{\itshape}
                        {0pt}{\pg@font}{.}{ }{}
 \theoremstyle{paper}
 \def\proofname{\normalfont\pg@font Proof}
}
\makeatother

\newtheorem{prop}{Proposition}[section]

\newcommand\egaldef{\stackrel{\mbox{\upshape\tiny def}}{=}}
\newcommand{\1}{\leavevmode\hbox{\rm \small1\kern-0.35em\normalsize1}}
\newcommand{\ind}[1]{\1_{\{#1\}}}

\begin{document}

\ifRR
  \RRetitle{Belief Propagation and Bethe approximation for Traffic Prediction}

  \RRtitle{Propagation de croyances et approximation de Bethe pour la
  prédiction de trafic} 
  \motcle{propagation de croyances, approximation de Bethe,
  reconstruction de trafic, prédiction, systèmes de transport
  intelligent, véhicules traceurs}

  \RRauthor{Cyril Furtlehner%
  \thanks{INRIA Futurs -- LRI, Bat. 490, Université Paris-Sud -- 91405
  Orsay cedex (France)}%
\and Jean-Marc Lasgouttes%
  \thanks{INRIA Rocquencourt -- Domaine de Voluceau B.P.\ 105 -- 78153 Le Chesnay cedex (France)}%
\and Arnaud de La Fortelle%
\thanks{École des Mines de Paris -- CAOR research centre -- 
    60, boulevard Saint-Michel -- 75272 Paris cedex 06 (France)}\footnotemark[2]}

  \RRtheme{\THNum \THCog}
  \RRprojets{Imara et Tao}
  \URRocq
  \RRdate{Mars 2007} 
  \RRNo{6144}
  \RRresume{On définit et étudie un algorithme de reconstruction
    utilisant l'algorithme « Belief Propagation » (propagation de
    croyances, BP) et l'approximation de Bethe. L'idée est d'encoder
    dans un graphe des données \emph{a priori} composées de
    corrélations ou de lois marginales et d'utiliser une procédure de
    passage de messages pour estimer l'état réel à partir
    d'informations temps-réel. Cette méthode, développée pour des
    besoins de prédiction de trafic, est particulièrement adaptée au
    cas où la seule information disponible provient de véhicules sonde
    (Floating Car Data). Nous proposons une discrétisation binaire du
    trafic s'appuyant sur le modèle d'Ising de physique statistique,
    permettant de reconstruire et de prédire le trafic en temps réel.
    Des propriétés générales de l'algorithme BP sont discutées dans ce
    contexte. En particulier une étude détaillée des propriétés de
    stabilité fonction des données \emph{a priori} et de la topologie
    du graphe est fournie.  Une étude numérique sur un modèle de
    trafic simplifié permet d'illustrer le fonctionnement de
    l'algorithme.  La façon de généraliser cette approche pour encoder
    une superposition de plusieurs états de trafic est discutée.  }
  
  \RRabstract{We define and study an inference algorithm based on
    ``belief propagation'' (BP) and the Bethe approximation.  The idea
    is to encode into a graph an a priori information composed of
    correlations or marginal probabilities of variables, and to use a
    message passing procedure to estimate the actual state from some
    extra real-time information.  This method is originally designed
    for traffic prediction and is particularly suitable in settings
    where the only information available is floating car data. We
    propose a discretized traffic description, based on the Ising
    model of statistical physics, in order to both reconstruct and
    predict the traffic in real time. General properties of BP are
    addressed in this context.  In particular, a detailed study of
    stability is proposed with respect to the a priori data and the
    graph topology.  The behavior of the algorithm is illustrated by
    numerical studies on a simple traffic toy model.  How this
    approach can be generalized to encode superposition of many
    traffic patterns is discussed.}

  \makeRR
\else
  \title{Belief Propagation and Bethe approximation for Traffic Prediction}

  \author{Cyril Furtlehner
  \thanks{INRIA Futurs -- LRI, Bat. 490, Université Paris-Sud -- 91405 Orsay Cedex}
\and Jean-Marc Lasgouttes
  \thanks{INRIA Rocquencourt -- Domaine de Voluceau B.P.\ 105 -- 78153 Le Chesnay Cedex (France)}
\and Arnaud de La Fortelle
\thanks{École des Mines de Paris -- CAOR research centre -- 
    60, boulevard Saint-Michel -- 75272 Paris cedex 06}\ \footnotemark[2]}
  \date{}
  \maketitle
  \begin{abstract}
    We define and study an inference algorithm based on ``belief
    propagation'' (BP) and the Bethe approximation.  The idea is to
    encode into a graph an a priori information composed of
    correlations or marginal probabilities of variables, and to use a
    message passing procedure to estimate the actual state from some
    extra real-time information.  This method is originally designed
    for traffic prediction and is particularly suitable in settings
    where the only information available is floating car data. We
    propose a discretized traffic description, based on the Ising
    model of statistical physics, in order to both reconstruct and
    predict the traffic in real time. General properties of BP are
    addressed in this context.  In particular, a detailed study of
    stability is proposed with respect to the a priori data and the
    graph topology.  The behavior of the algorithm is illustrated by
    numerical studies on a simple traffic toy model.  How this
    approach can be generalized to encode superposition of many
    traffic patterns is discussed.
  \end{abstract}


\fi

\section{Introduction}

With an estimated $1$\% GDP cost in the European Union (i.e.\ more
than hundred billions euros), congestion is not only a time waste for
drivers and an environmental challenge, but also an economic issue.
This is why the European commission financed the REACT project, where
new traffic prediction models have been developed. These predictions
are to be used to inform the public and possibly to regulate the
traffic.

Today, some urban and inter-urban areas have
traffic management and advice systems that collect data from
stationary sensors, analyze them, and post notices about road
conditions ahead and recommended speed limits on display signs located
at various points along specific routes. However, these systems are
not available everywhere and they are virtually non-existent on rural
areas. With rural road crashes accounting for more than $60$\% of all
road fatalities in OECD (Organization for Economic Cooperation and
Development) countries, the need for a system that can cover these
roads is compelling if a significant reduction in traffic deaths is to
be achievable.

The REACT project combines a traditional traffic prediction approach
on equipped motorways with an innovative approach on non-equipped
roads. The idea is to obtain floating car data from a fleet of probe
vehicles and reconstruct the traffic conditions from this partial
information. To understand why it is not possible to fuse these two
parts, we have to go a bit more into prediction algorithms details.

Two types of approaches are usually distinguished, namely \emph{data
  driven} (application of statistical models to a large amount of
data, for example regression analysis) and \emph{model based}
(simulation or mathematical models explaining the traffic patterns).
As we stated before, the choice is largely led by the availability of
data. In our case, since little data is available on non-equipped roads
(only the equipped vehicles driving along the observed roads), the
model driven approach is the only feasible one. For more information
about traffic prediction methods, we refer the reader
to~\cite{Benz,Prime,VT}.

Most current traffic models are deterministic, described either at a
macroscopic level by a set of differential equations linking variables
such as flow and density, or by Newton's law at a microscopic level
where each individual car is considered. Intermediate descriptions are
essentially kinetic models, like for example cellular
automata~\cite{NaSch}, which are very well adapted to freeway traffic
modeling and adapted to some extent to urban traffic
modeling~\cite{EssSch}. Traffic flow models are quite adapted and
efficient on motorways where fluid approximation of the traffic is
reasonable; they tend to fail for cities or rural roads. The reason is
that the velocity flow field is subject to much greater fluctuations
induced by the nature of the network (presence of intersections and
short distance between two intersections) than by the traffic itself.
These fluctuations are both spatial and temporal (a red or green
traffic light at a cross-road, a road-work, etc). There is no local
stationary regime for the velocity, the dynamics are dominated by the
fluctuations. 

We propose in this paper a hybrid approach in the continuation
of~\cite{FuLa}, by taking full advantage of the statistical nature of
the information, in combination with a stochastic modeling of traffic
patterns. In order to reconstruct the traffic and make predictions, we
propose a model---the Bethe approximation (BA)---to encode the
statistical fluctuations and stochastic evolution of the traffic and
an algorithm---the belief propagation (BP) algorithm---to decode the
information. Those concepts are familiar to the computer science and
statistical physics communities since it was shown~\cite{YeFrWe3} that
the output of BP is in general the Bethe approximation~\cite{Bethe}.

The paper is organized as follows: Section~\ref{traffdes} describes
the model and its relationship to the Ising model and the Bethe
approximation. The inference problem and our strategy to tackle it
using the Belief Propagation algorithm are stated in
Section~\ref{algorithmes}. The implementation of these ideas requires
some new results about the BP algorithm, which are the subject of
Section~\ref{sec:g_properties}; this concerns in particular the effect
of the normalization of the messages, the parameterization of the
model and the stability of the fixed points. Section~\ref{simulation}
is devoted to implementation details of the decoding algorithm and to some
numerical results illustrating the method. Finally, some new research
directions are proposed in Section~\ref{sec:conclusion}.

\section{Traffic description and statistical physics}\label{traffdes}

The graph onto which we apply the belief propagation procedure is made
of space-time vertices that encode both a location (road link) and a
time (discretized on a few minutes scale). More precisely, the set of
vertices is $\V=\mathcal{L}\otimes \mathbb{Z}^+$, where $\mathcal{L}$
corresponds to the links of the network and $\mathbb{Z}^+$ to the time
discretization. To each point $\alpha=(\ell,t)\in\V$, we attach an
information $\tau_\alpha\in\{0,1\}$ indicating the state of the
traffic ($1$ if congested, $0$ otherwise). Each cell is correlated to
its neighbors (in time and space) and the evaluation of this local
correlation determines the model. In other words, we assume that the
joint probability distribution of $\tau_\V\egaldef\{\tau_\alpha,\
\alpha\in \V\}\in\{0,1\}^\V$ is of the form
\begin{equation}\label{eq:joint}
p(\{\tau_\alpha, \alpha\in\V\}) \quad
= \prod_{\alpha\in\V}\phi_\alpha(\tau_\alpha)
  \prod_{(\alpha,\beta)\in \E}\psi_{\alpha\beta}(\tau_\alpha,\tau_\beta)
\end{equation}
where $\E\subset\V^2$ is the set of edges, and the local
correlations are encoded in the functions $\psi$ and $\phi$. $\V$
together with $\E$ describe the space-time graph $\G$ and
$\V(\alpha)\subset\V$ denotes the set of neighbors of vertex $\alpha$.

\begin{figure}
\centering
\includegraphics[width=0.7\textwidth]{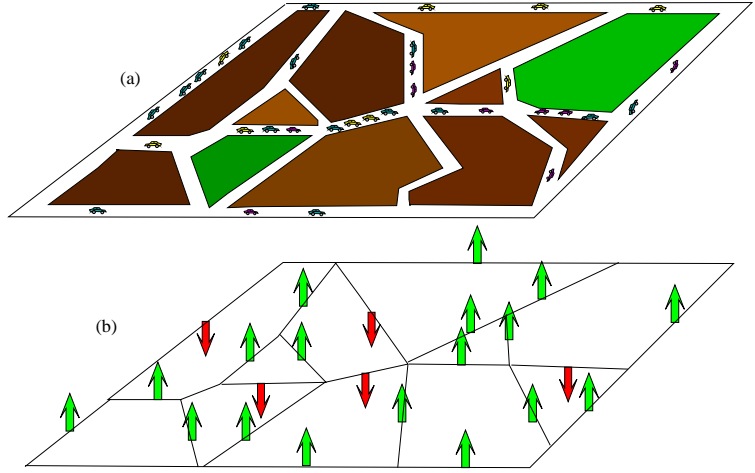}
\caption{\label{graph1} Traffic network (a) and Ising model (b) on a random graph}
\end{figure}

The model described by~(\ref{eq:joint}) is actually equivalent to an
Ising model~\cite{Ising} on $\G$, with arbitrary coupling between 
adjacent spins, the up or down orientation of each spin indicating the
status of the corresponding link (Figure~\ref{graph1}).

The homogeneous Ising model (uniform coupling constants) is a
well-studied model of ferro (positive coupling) or anti-ferro
(negative coupling) material in statistical physics. It displays a
phase transition phenomenon with respect to the value of the coupling.
At weak coupling, only one disordered state occurs, where spins are
randomly distributed around a mean-zero value. Conversely, when the
coupling is strong, there are two equally probable states that
correspond to the onset of a macroscopic magnetization either in the
up or down direction: each spin has a larger probability to be
oriented in the privileged direction than in the opposite one.

From the point of view of a traffic network, this means that such a
model is able to describe three possible traffic regimes: fluid (most
of the spins up), congested (most of the spins down) and dense
(roughly half of the links are congested). For real situations, we
expect other types of congestion patterns, and we seek to associate
them either to the $p$-state Potts model if we extend the binary to $p$-ary description,
or to the possible states of an inhomogeneous Ising model with
frustration (i.e.\ 
with possibly negative coupling parameters), referred as spin glasses in 
statistical physics~\cite{MePaVi}.
When such a system is frustrated because some negative couplings,
leading to a certain number of contradictions, a proliferation of
meta-stable states occurs, which eventually scales exponentially with
the size of the system.

On a simply connected graph, the knowledge of the one-vertex and
two-vertices marginal probabilities is sufficient~\cite{Pearl} to fully
determine the measure (\ref{eq:joint}).
\begin{equation}\label{eq:tree}
p(\tau_\V) = \frac{\prod_{(\alpha,\beta)\in \E} 
\hat p_{\alpha\beta}(\tau_\alpha,\tau_\beta)}{\prod_{\alpha\in \V}
\hat p(\tau_\alpha)^{q_\alpha-1}}
= \prod_{\alpha\in\V}\hat p_\alpha(\tau_\alpha)
  \prod_{(\alpha,\beta)\in \E} 
         \frac{\hat p_{\alpha\beta}(\tau_\alpha,\tau_\beta)} 
     {\hat p_{\alpha}(\tau_\alpha)\hat p_{\beta}(\tau_\beta)},
\end{equation}
where $q_\alpha$ denotes the number of neighbors of $\alpha$. Since
our space time graph $\G$ is multi-connected, this relationship
between local marginals and the full joint probability measure can
only be an approximation, which in the context of statistical physics
is referred to as the Bethe approximation. This approximation is
provided by the minimum of the so-called Bethe free energy, which,
based on the form~(\ref{eq:tree}), is an approximate form of the
Kullback-Leibler distance,
\[
D(b(\tau_\V)\|p(\tau_\V)) \egaldef \sum_{\tau_\V}b(\tau_\V)\ln \frac{b(\tau_\V)}
{p(\tau_\V)},
\]
and which rewrites in terms of a free energy as
\[
D(b(\tau_\V)\|p(\tau_\V)) = \mathcal{F}(b(\tau_\V)) - \mathcal{F}(p(\tau_\V)),
\]
where
\begin{equation}\label{eq:bethe}
\mathcal{F}(b(\tau_\V)) \egaldef \mathcal{U}(b(\tau_\V))-\mathcal{S}(b(\tau_\V)),
\end{equation}
with the respective definitions of the energy $\mathcal{U}$ and of the
entropy $\mathcal{S}$
\begin{align*}
\mathcal{U}(b(\tau_\V)) 
 &\egaldef -\sum_{(\alpha,\beta)\in \E} b_{\alpha\beta}(\tau_\alpha,\tau_\beta)
                      \log \psi_{\alpha\beta}(\tau_\alpha,\tau_\beta) 
- \sum_{\alpha\in \V} b_{\alpha}(\tau_\alpha) \log \phi_{\alpha}(\tau_\alpha),\\
\mathcal{S}(b(\tau_\V)) 
 &\egaldef -\sum_{(\alpha,\beta)\in \E} b_{\alpha\beta}(\tau_\alpha,\tau_\beta)
                      \log b_{\alpha\beta}(\tau_\alpha,\tau_\beta) 
- \sum_{\alpha\in \V} b_{\alpha}(\tau_\alpha) \log b_{\alpha}(\tau_\alpha).
\end{align*}

In practice, what we retain from an Ising description is the
possibility to encode a certain number of traffic patterns in a
statistical physics model. This property is actually shared also by
the Bethe Approximation (BA) and this is the reason for us to
directly encode the traffic patterns in a BA rather than the
inhomogeneous Ising model itself, based on historical data, and to
avoid therefore an intermediate approximation step. BA simply provides
us with a set of marginals probabilities that we try to match with the
historical data. But this set, which is the result of an iterative
procedure, is not necessarily unique (see for example~\cite{MoKa}) and
the proliferation of possible solutions depends on the frustration
induced by the historical correlations used to define the $\psi$'s of
(\ref{eq:joint}). The setting of our model consists therefore into an
optimization procedure of the matching between the set of historical
values obtained from probe vehicles and the set of marginal
probabilities of the BA.

\medskip The data collected from the probe vehicles is used in two
different ways. The most evident one is that the data of the current
day directly influences the prediction. In parallel, this data is
collected over long periods (weeks or months) in order to estimate the
model~(\ref{eq:joint}). Typical historical data that is
accumulated is
\begin{itemize}
\item $\hat p_\alpha(\tau_\alpha)$: the probability that vertex $\alpha$ is
  congested ($\tau_\alpha=1$) or not ($\tau_\alpha=0$);
\item $\hat p_{\alpha\beta}(\tau_\alpha,\tau_\beta)$: the probability
  that a probe vehicle going from $\alpha$ to $\beta\in\V(\alpha)$ 
  finds $\alpha$ with state $\tau_\alpha$ and $\beta$ with state $\tau_\beta$.
\end{itemize}
The edges $(\alpha,\beta)$ of the space time graph $\G$ are constructed 
based on the presence of a measured mutual information between $\alpha$ 
and $\beta$, which is the case when
$\hat p_{\alpha\beta}(\tau_\alpha,\tau_\beta) \ne
\hat p_\alpha(\tau_\alpha)\hat p_\beta(\tau_\beta)$. 

\section{The reconstruction and prediction algorithm}\label{algorithmes}

\subsection{Statement of the inference problem}
We turn now to our present work concerning an inference problem, which
we set in general terms as follows: a set of observables $\tau_\V =
\{\tau_\alpha,\ \alpha\in \V\}$, which are stochastic variables are
attached to the set $\V$ of vertices of a graph. For each edge
$(\alpha,\beta)\in \E$ of the graph, an accumulation of repetitive
observations allows to build the empirical marginal probabilities
$\{\hat p_{\alpha\beta}\}$. The question is then: given the values of
a subset $\tau_{\V^*} = \{\tau_{\alpha^*},\ \alpha^*\in \V^*\}$, what
prediction can be made concerning $\overline{\V^*}$, the complementary
set of $\V^*$ in $\V$?

There are two main  issues:
\begin{itemize}
\item how to encode the historical observations (inverse problem) in
  an Ising model, such that its marginal probabilities on the edges
  coincide with the $\hat p_{\alpha\beta}$?
\item how to decode in the most efficient manner---typically in real
  time---this information, in terms of conditional probabilities
  $P(\tau_\alpha|\tau_{\V^*})$?
\end{itemize}

The answer to the second question will somehow give a hint to the first one.

\subsection{The belief propagation algorithm}

BP is a message passing procedure, which output is a set of estimated
marginal probabilities, the beliefs $b_{\alpha\beta}$~\cite{Pearl}.
The idea of the BP algorithm is to factor the marginal probability at
a given site in a product of contributions coming from neighboring
sites, which are the messages. The messages sent by a vertex $\alpha$ to
$\beta\in \V(\alpha)$ depends on the messages it received previously
from other vertices:
\begin{equation}\label{urules}
m_{\alpha\to\beta}(\tau_{\beta})
\gets
\sum_{\tau_\alpha\in\{0,1\}} n_{\alpha\to\beta}(\tau_\alpha) 
\phi_\alpha(\tau_\alpha)\psi_{\alpha\beta}(\tau_\alpha, \tau_\beta),
\end{equation}
where
\begin{equation}\label{urulesn}
n_{\alpha\to\beta}(\tau_\alpha)\egaldef \prod_{\gamma\in \V(\alpha)\setminus\{\beta\}}
m_{\gamma\to\alpha}(\tau_\alpha).
\end{equation}
In practice, the messages will be normalized so that 
\begin{equation}\label{eq:normalization}
 \sum_{\tau_\beta\in\{0,1\}} m_{\alpha\to\beta}(\tau_{\beta})=1.
\end{equation}
We will come back to the effects of this in Section~\ref{sec:normalization}.

The output of the algorithm is a set of \emph{beliefs}, which are an
approximation of the one-vertex and two-vertices marginals of $p(\tau_\V)$.
The beliefs $b_\alpha$ are reconstructed according to
\begin{equation}\label{belief1}
b_\alpha(\tau_\alpha) \propto 
\phi_\alpha(\tau_\alpha)\prod_{\beta\in \V(\alpha)} m_{\beta\to\alpha}(\tau_\alpha),
\end{equation}
and, similarly, the belief $b_{\alpha\beta}$ of the joint probability of
$(\tau_\alpha,\tau_\beta)$ is given by
\begin{equation}\label{belief2}
 b_{\alpha\beta}(\tau_\alpha,\tau_\beta) 
\propto n_{\alpha\to\beta}(\tau_\alpha)n_{\beta\to\alpha}(\tau_\beta)
\times\phi_\alpha(\tau_\alpha)\phi_\beta(\tau_\beta)
\psi_{\alpha\beta}(\tau_\alpha, \tau_\beta).
\end{equation}
In the formulas above and in the remainder of this paper, the
proportionality symbol $\propto$ indicates that one must normalize the
beliefs so that they sum to $1$.

A simple computation shows that equations~(\ref{belief1}) and
(\ref{belief2}) are compatible, since
(\ref{urules})--(\ref{urulesn}) imply that
\[
  \sum_{\tau_\alpha\in\{0,1\}} b_{\alpha\beta}(\tau_\alpha,\tau_\beta)
  = b_\beta(\tau_\beta).
\]

It has been realized a few years ago~\cite{YeFrWe} that the fixed
points of the BP algorithm coincide with local minima of the Bethe
free energy~(\ref{eq:bethe}). This justifies that we can use this
algorithm to approximate our Ising model.

\medskip We propose to use the BP algorithm for two purposes:
estimation of the model parameters (the functions $\phi$ and $\psi$) from
historical data and reconstruction of traffic from current data.

\subsection{Setting the model with Belief Propagation}\label{heuristic}
The fixed points of the BP algorithm (and therefore the Bethe
approximation) allow to approximate the joint marginal probability
$p_{\alpha\beta}$ when the functions $\psi_{\alpha\beta}$ and
$\phi_\alpha$ are known. Conversely, it can provide good candidates
for $\psi_{\alpha\beta}$ and $\phi_\alpha$ from the historical values
$\hat p_{\alpha\beta}$ and $\hat p_\alpha$.

To set up our model, we are looking for a fixed point of the BP
algorithm satisfying (\ref{urules})--(\ref{urulesn}) and such that
$b_{\alpha\beta}(\tau_\alpha,\tau_\beta) = \hat
p_{\alpha\beta}(\tau_\alpha,\tau_\beta)$ and therefore
$b_{\alpha}(\tau_\alpha)=\hat p_{\alpha}(\tau_\alpha)$.


It is easy to check that the following choice of $\phi$ and $\psi$, 
\begin{eqnarray}
  \psi_{\alpha\beta}(\tau_\alpha, \tau_\beta) 
  &=&\frac{\hat p_{\alpha\beta}(\tau_\alpha,\tau_\beta)} 
     {\hat p_{\alpha}(\tau_\alpha)\hat p_{\beta}(\tau_\beta)}\label{bethepsi},\\
  \phi_\alpha(\tau_\alpha) &=& \hat p_{\alpha}(\tau_\alpha)\label{bethephi},
\end{eqnarray}
leads (\ref{eq:joint}) to coincide with (\ref{eq:tree}). They
correspond to a normalized BP fixed point for which all messages are
equal to $1/2$. There is however no guarantee that this fixed point is
a stable fixed point; actually, for an Ising-type system below the
critical temperature, we often observe that this point is unstable
(see Section~\ref{simulation} for the simulation results). It will be
shown however in Section~\ref{sec:fixedpoints} that this form of
$\phi$ and $\psi$ is in some sense canonical.

\subsection{Traffic reconstruction and prediction}
Let $\V^*$ be the set of vertices that have been visited by probe
vehicles. Reconstructing traffic from the data gathered
by those vehicles is equivalent to evaluating the conditional probability
\[
 p_{\alpha}(\tau_\alpha|\tau_{\V^*}) 
  =  \frac{p_{\alpha,\V^*}(\tau_\alpha,\tau_{\V^*})}
          {p_{\V^*}(\tau_{\V^*})},
\]
where $\tau_{\V^*}$ is a shorthand notation for the set 
$\{\tau_{\alpha^*}\}_{\alpha^*\in\V^*}$.

The BP algorithm applies to this case if a specific rule is defined
for vertices $\alpha^*\in\V^*$: since the value of $\tau_{\alpha^*}$ is
known, there is no need to sum over possible values and~(\ref{urules}) becomes
\[
m_{\alpha^*\to\beta}(\tau_{\beta})
\gets n_{\alpha^*\to\beta}(\tau_{\alpha^*})
\phi_{\alpha^*}(\tau_{\alpha^*})\psi_{\alpha^*\beta}(\tau_{\alpha^*}, \tau_\beta).
\]

The resulting algorithm is supposed to be run in real time, over a
graph which corresponds to a time window (typically a few hours)
centered around present time, with probe vehicle data added as it is
available. In this perspective, the reconstruction and prediction
operations are done simultaneously on an equal footing, the
distinction being simply the time-stamp (past for reconstruction or
future for prediction) of a given computed belief.  The output of the
previous run can be used as initial messages for a new run, in order
to speedup convergence. Full re-initialization (typically a random set
of initial messages) has to be performed within a time interval of the
order but smaller than the time-scale of typical traffic fluctuations.

\section{Some general properties of the Belief Propagation algorithm}\label{sec:g_properties}

This section contains several theoretical results on the BP algorithm.
Although they are stated in the context of Section~\ref{traffdes},
most of these results can be trivially extended to a general factor
graph and variables taking more than two values (transforming the
Ising model into a Potts model), except possibly for
Section~\ref{sec:stability}.

\subsection{Building the model from its fixed points}\label{sec:fixedpoints}

The particular use that we make of the Bethe approximation, as
outlined in Section~\ref{heuristic}, means that the output of the
algorithm takes precedence over the underlying Ising model, which is
an unusual situation. The following propositions shows how to estimate
$\phi_\alpha$ and $\psi_{\alpha\beta}$ from the historical values
$\hat p_\alpha$ and $\hat p_{\alpha\beta}$.

Let us start with a direct consequence of the BP fixed point
equations. The following straightforward proposition
extends~(\ref{eq:tree}) to the case of a non-tree structure.
\begin{prop}
  A set of beliefs $\{b_\alpha,b_{\alpha\beta}\}$ corresponding to a
  BP fixed point of (\ref{urules})--(\ref{belief2}) always satisfies
\[
p(\tau_\V) = \frac{\prod_{\alpha,\beta}b_{\alpha\beta}(\tau_\alpha,\tau_\beta)}
{\prod_\alpha b_\alpha^{q_\alpha-1}(\tau_\alpha)}
=\prod_{\alpha\in\V}b_\alpha(\tau_\alpha)
 \prod_{(\alpha,\beta)\in \E} 
         \frac{b_{\alpha\beta}(\tau_\alpha,\tau_\beta)} 
     {b_{\alpha}(\tau_\alpha)b_{\beta}(\tau_\beta)}.
\] 
\end{prop}

\begin{proof} This is a simple consequence
  of~(\ref{belief1})--(\ref{belief2}).
\end{proof}

What this proposition means is that different BP fixed points
correspond to different factorizations of the joint
measure~(\ref{eq:joint}). The knowledge of a set of beliefs
is thus sufficient to determine the underlying Ising model and
consequently the other fixed points of the algorithm.

The following proposition gives more insight on how the different
components of~(\ref{eq:joint}) can be written in terms of the BP fixed
points.

\begin{prop}\label{proppsi}
Assume that there exists a fixed point of the BP algorithm satisfying
(\ref{urules})--(\ref{belief2}) and such that
\begin{equation}
b_{\alpha\beta}(\tau_\alpha,\tau_\beta)=\hat
p_{\alpha\beta}(\tau_\alpha,\tau_\beta),\text{ and therefore }
b_{\alpha}(\tau_\alpha)=\hat p_{\alpha}(\tau_\alpha).\label{goodbelief}
\end{equation} 
Then the following equalities hold
\begin{eqnarray}
\psi_{\alpha\beta}(\tau_\alpha, \tau_\beta) 
&=& \frac{\hat p_{\alpha\beta}(\tau_\alpha,\tau_\beta)} 
     {\hat p_{\alpha}(\tau_\alpha)\hat p_{\beta}(\tau_\beta)}
m_{\alpha\to\beta}(\tau_\beta)m_{\beta\to\alpha}(\tau_\alpha),\label{psi}\\
\phi_\alpha(\tau_\alpha)
&=& \frac{\hat p_\alpha(\tau_\alpha)}
         {\prod_{\beta\in \V(\alpha)}m_{\beta\to\alpha}(\tau_\alpha)}.\label{phi}
\end{eqnarray}

Conversely, assume that there exist boolean functions
$\mu_{\alpha\beta}(\tau_\beta)$ such that
\begin{eqnarray}
\psi_{\alpha\beta}(\tau_\alpha, \tau_\beta) 
&=& \frac{\hat p_{\alpha\beta}(\tau_\alpha,\tau_\beta)} 
     {\hat p_{\alpha}(\tau_\alpha)\hat p_{\beta}(\tau_\beta)}
    \mu_{\alpha\beta}(\tau_\beta)\mu_{\beta\alpha}(\tau_\alpha),\label{psimu}\\
\phi_\alpha(\tau_\alpha)
&=& \frac{\hat p_\alpha(\tau_\alpha)}
         {\prod_{\beta\in \V(\alpha)} \mu_{\beta\alpha}(\tau_\alpha) }.\label{bmu}
\end{eqnarray}
Then $m_{\alpha\to\beta}=\mu_{\alpha\beta}$ is a fixed point of the BP
algorithm and~(\ref{goodbelief}) holds.
\end{prop}

\begin{proof}
Relation (\ref{psi}) is obtained by rewriting~(\ref{belief1})
and~(\ref{belief2}) as
\begin{eqnarray}
\psi_{\alpha\beta}(\tau_\alpha, \tau_\beta) 
&=&\frac{b_{\alpha\beta}(\tau_\alpha,\tau_\beta)} 
     {\phi_\alpha(\tau_\alpha)n_{\alpha\to\beta}(\tau_\alpha)
n_{\beta\to\alpha}(\tau_\beta)\phi_\beta(\tau_\beta)}\nonumber\\
&=& \frac{b_{\alpha\beta}(\tau_\alpha,\tau_\beta)} 
     {b_{\alpha}(\tau_\alpha)b_{\beta}(\tau_\beta)}
m_{\alpha\to\beta}(\tau_\beta)m_{\beta\to\alpha}(\tau_\alpha).\label{psi1}
\end{eqnarray}

To prove the second assertion, the first step is to show that
$\mu_{\alpha\beta}$ is a BP fixed point:
\begin{eqnarray*}
\lefteqn{\sum_{\tau_\alpha\in\{0,1\}} 
  \biggl[\prod_{\gamma\in \V(\alpha)\setminus\{\beta\}}
  \mu_{\gamma\alpha}(\tau_\alpha)\biggr]
  \phi_\alpha(\tau_\alpha)\psi_{\alpha\beta}(\tau_\alpha, \tau_\beta)}\qquad\qquad\\
&=&\sum_{\tau_\alpha\in\{0,1\}} 
   \frac{\hat p_\alpha(\tau_\alpha)}{\mu_{\beta\alpha}(\tau_\alpha)}
\frac{\hat p_{\alpha\beta}(\tau_\alpha,\tau_\beta)} 
     {\hat p_{\alpha}(\tau_\alpha)\hat p_{\beta}(\tau_\beta)}
    \mu_{\alpha\beta}(\tau_\beta)\mu_{\beta\alpha}(\tau_\alpha)\\
&=& \sum_{\tau_\alpha\in\{0,1\}} \frac{\hat
  p_{\alpha\beta}(\tau_\alpha,\tau_\beta)}{\hat p_{\beta}(\tau_\beta)}
  \mu_{\alpha\beta}(\tau_\beta)\\
&=& \mu_{\alpha\beta}(\tau_\beta).
\end{eqnarray*}
For this fixed point, (\ref{belief2}) reduces to (\ref{goodbelief}),
which concludes the proof of the proposition.
\end{proof}

While Proposition~\ref{proppsi} seems to indicate that there is some
leeway in choosing $\psi_{\alpha\beta}$, a proper change of variables
shows that all the choices are equivalent. Let us define the following
new set of messages
\[
  x_{\alpha\to\beta}(\tau_\beta)\egaldef
  \frac{m_{\alpha\to\beta}(\tau_\beta)}{\mu_{\alpha\beta}(\tau_\beta)}.
\]

Equation~(\ref{urules}) then becomes
\begin{eqnarray*}
\lefteqn{x_{\alpha\to\beta}(\tau_\beta)\mu_{\alpha\beta}(\tau_\beta)}\quad\\
&=& \sum_{\tau_\alpha\in\{0,1\}} \biggl[\prod_{\gamma\in \V(\alpha)\setminus\{\beta\}}
x_{\gamma\to\alpha}(\tau_\alpha)\mu_{\gamma\alpha}(\tau_\alpha)\biggr]
\phi_\alpha(\tau_\alpha)\psi_{\alpha\beta}(\tau_\alpha, \tau_\beta)\\
&=&\sum_{\tau_\alpha\in\{0,1\}}\biggl[\prod_{\gamma\in
  \V(\alpha)\setminus\{\beta\}}x_{\gamma\to\alpha}(\tau_\alpha)\biggr]
  \frac{\hat p_{\alpha}(\tau_\alpha)}{\mu_{\beta\alpha}(\tau_\alpha)}
    \frac{\hat p_{\alpha\beta}(\tau_\alpha,\tau_\beta)} 
     {\hat p_{\alpha}(\tau_\alpha)\hat p_{\beta}(\tau_\beta)}
    \mu_{\alpha\beta}(\tau_\beta)\mu_{\beta\alpha}(\tau_\alpha)\\
&=& \sum_{\tau_\alpha\in\{0,1\}}\biggl[\prod_{\gamma\in
  \V(\alpha)\setminus\{\beta\}}x_{\gamma\to\alpha}(\tau_\alpha)\biggr]
    \frac{\hat p_{\alpha\beta}(\tau_\alpha,\tau_\beta)} 
     {\hat p_{\beta}(\tau_\beta)}
    \mu_{\alpha\beta}(\tau_\beta),
\end{eqnarray*}
and therefore
\[
x_{\alpha\to\beta}(\tau_\beta)
=  \sum_{\tau_\alpha\in\{0,1\}}\biggl[\prod_{\gamma\in
  \V(\alpha)\setminus\{\beta\}}x_{\gamma\to\alpha}(\tau_\alpha)\biggr]
    \frac{\hat p_{\alpha\beta}(\tau_\alpha,\tau_\beta)} 
     {\hat p_{\beta}(\tau_\beta)}.
\]

This version of the BP algorithm is thus equivalent to
the heuristic choice~(\ref{bethepsi})--(\ref{bethephi}), which
corresponds to the trivial fixed point
$x_{\alpha\to\beta}(\tau_\beta)\equiv 1$.

Since it is equivalent in terms of convergence to the original choice
of $\psi_{\alpha\beta}$ and $\phi_\alpha$, this can be seen as the
canonical choice of functions to define our Ising model.

The freedom we have in the definition of $\phi$ and $\psi$ yields the
following possibility:
\begin{prop}\label{refchange}
  Assume that the schema (\ref{urules})--(\ref{urulesn}) admits a set
  $\{m^i\}$, $i\in\mathcal{I}$, of fixed points with corresponding
  beliefs $\{b^i\}$. For any $i_0\in\mathcal{I}$, choosing  $i_0$ as a
  reference state by changing $\phi$ and $\psi$ according to
\begin{align*}
\psi_{\alpha\beta}^{i_0}(\tau_\alpha, \tau_\beta) 
&= \frac{b_{\alpha\beta}^{i_0}(\tau_\alpha,\tau_\beta)} 
     {b_{\alpha}^{i_0}(\tau_\alpha) b_{\beta}^{i_0}(\tau_\beta)},\\[0.2cm]
\phi_\alpha^{i_0}(\tau_\alpha)
&= b^{i_0}_\alpha(\tau_\alpha),
\end{align*}
yields a new BP scheme, with unchanged beliefs $\{b^i\}$, but
with a new set of fixed points
\[
m_{\alpha\to\beta}^{(i/i_0)}(\tau_\beta) = \frac{m_{\alpha\to\beta}^{(i)}(\tau_\beta)}
{m_{\alpha\to\beta}^{(i_0)}(\tau_\beta)}. 
\] 
In particular, the new reference fixed point $\{m^{(i_0/i_0)}\}$ has all
its components identically equal to $1$.
\end{prop}

\subsection{Normalization and fixed points}\label{sec:normalization}
We discuss here a feature of the algorithm which did not get that much
attention in the literature, which is the possibility of normalizing
the messages and its consequences on the results. In most studies, it
is assumed that the messages are normalized so that
(\ref{eq:normalization}) holds. The update rule~(\ref{urules}) indeed indicates
that there is an important risk to see the messages converge to $0$ or
diverge to infinity. It is however not immediate to check that the
normalized version of the algorithm has the same fixed points as the
original one (and therefore the Bethe approximation).

In order to make the definition of normalization clear, define the mapping
\[
 \Theta_{\alpha\beta}(m)(\tau_\beta)\egaldef
  \sum_{\tau_\alpha\in\{0,1\}} \biggl[\prod_{\gamma\in \V(\alpha)\setminus\{\beta\}}
m_{\gamma\to\alpha}(\tau_\alpha)\biggr]
\phi_\alpha(\tau_\alpha)\psi_{\alpha\beta}(\tau_\alpha, \tau_\beta),
\]

Then the normalized version of BP is defined by the following update
rule
\begin{equation}\label{eq:normrule}
  \tilde m_{\alpha\to\beta}(\tau_\beta)
\gets\frac{\Theta_{\alpha\beta}(\tilde m)(\tau_\beta)}
       {\Theta_{\alpha\beta}(\tilde m)(0)
        +\Theta_{\alpha\beta}(\tilde m)(1)}.
\end{equation}

The relation between the fixed points of BP and normalized BP can be
described as follows.

\begin{prop}\label{prop:normalization}
  Any normalized fixed point (except $0$) of the BP
  algorithm is a fixed point of the version of BP algorithm with normalized
  messages.
  
  Conversely, a fixed point of the BP algorithm with normalized
  messages corresponds (through multiplication by a proper constant)
  to an unique fixed point of the basic BP algorithm, except possibly
  when the graph $\G$ has exactly one cycle.
\end{prop}

\begin{proof}
Let $m$ be a non-null fixed point of the BP algorithm, that is
\[
  m_{\alpha\to\beta}(\tau_\beta)=\Theta_{\alpha\beta}(m)(\tau_\beta),\qquad\forall(\alpha,\beta)\in\E
\]
and let
\[
 \tilde m_{\alpha\to\beta}(\tau_\beta)
  =\frac{m_{\alpha\to\beta}(\tau_\beta)}{m_{\alpha\to\beta}(0)+m_{\alpha\to\beta}(1)}
  = k_{\alpha\beta}\,m_{\alpha\to\beta}(\tau_\beta).
\]
From its definition, $\Theta_{\alpha\beta}$ is multilinear and
\[
  \Theta_{\alpha\beta}(\tilde m)(\tau_\beta)=\biggl[\prod_{\gamma\in
  \V(\alpha)\setminus\{\beta\}}k_{\gamma\alpha}\biggr]\Theta_{\alpha\beta}(m)(\tau_\beta),
\]
and therefore $\tilde m$ is a fixed point of the schema~(\ref{eq:normrule}).

Conversely, let $\tilde m$ be a fixed point of~(\ref{eq:normrule}).
Then there exists a set of constants $K_{\alpha\beta}$ satisfying
\[
 \Theta_{\alpha\beta}(\tilde m)(\tau_\beta) = K_{\alpha\beta}\,\tilde m_{\alpha\to\beta}(\tau_\beta).
\]
Let us find a set of constants $c_{\alpha\beta}$ such that
\[
m_{\alpha\to\beta}(\tau_\beta)=c_{\alpha\beta}\,\tilde
m_{\alpha\to\beta}(\tau_\beta),
\]
be a non-zero fixed point of~(\ref{urules}).
We have
\begin{align*}
 \Theta_{\alpha\beta}(m)(\tau_\beta) 
&= \biggl[\prod_{\gamma\in  \V(\alpha)\setminus\{\beta\}}
c_{\gamma\alpha}\biggr] \Theta_{\alpha\beta}(\tilde m)(\tau_\beta) \\
&= \biggl[\prod_{\gamma\in  \V(\alpha)\setminus\{\beta\}}
c_{\gamma\alpha}\biggr] K_{\alpha\beta}\,\tilde
m_{\alpha\to\beta}(\tau_\beta)\\
&= \frac{1}{c_{\alpha\beta}}
   \biggl[\prod_{\gamma\in  \V(\alpha)\setminus\{\beta\}}
c_{\gamma\alpha}\biggr] K_{\alpha\beta}\,m_{\alpha\to\beta}(\tau_\beta),
\end{align*}
and therefore
\begin{equation}\label{eq:eigen1}
\log c_{\alpha\beta} - \sum_{\gamma\in \V(\alpha)\setminus\{\beta\}}
\log c_{\gamma\alpha}=\log K_{\alpha\beta}.
\end{equation}
Solving this equation amounts to invert a matrix $I-A$
where $A$ is an incidence matrix on the dual factor graph (the graph which 
connects oriented pairs $(\alpha,\beta)\in\E$, see Figure~\ref{meta}). 
Let $v_{\alpha\beta} = \log c_{\alpha\beta}$.
The homogeneous  equation rewrites
\begin{equation}\label{eq:eigen2}
v_{\alpha\beta} + v_{\beta\alpha} = \sum_{\gamma\in \V(\alpha)}
v_{\gamma\alpha}.
\end{equation}
When a non-zero solution exists, then a
simple symmetry argument shows that the right-hand side does not
depend on either $\alpha$ or $\beta$ and therefore can be set to $1$
without loss of generality. Therefore, summing over all oriented edges,
\begin{align*}
2|\E| 
&= 2\sum_{(\alpha,\beta)\in \E}(v_{\alpha\beta} + v_{\beta\alpha})\\
&= \sum_{\alpha\in\V}\sum_{\beta\in \V(\alpha)}v_{\alpha\beta}
  +\sum_{\beta\in\V}\sum_{\alpha\in \V(\beta)}v_{\beta\alpha}\\
&= 2|\V|,
\end{align*}
with $|\E|$ and $|\V|$ respectively the number of edges and vertices.
Since $\G$ has only one component, by the well-known 
formula~\cite{Berge}, the number of cycles in the graph
is $|\E|-|\V|+1$, only graphs with one cycle give possibly rise to a
non-zero solution to~(\ref{eq:eigen2}). Conversely, when a graph has
one unique cycle, it is possible to provide an partial ordering of
vertices such that each vertex has exactly one neighbor greater than
itself, and $v_{\alpha\beta}=\ind{\alpha>\beta}$
is a solution to~(\ref{eq:eigen2}).
\end{proof}

This proposition does not describe what happens when $\G$ has exactly
one cycle. The existence of a solution to (\ref{eq:eigen1}) actually
depends on the value of $\log K$, which itself depends on the fixed
point $\tilde m$. However, since BP is known to converge in a finite
number of steps for graphs with at most $1$ cycle, normalization is
not useful in this situation.

From now on reference to the BP algorithm is to be understood as its
normalized version.

\subsection{Stability of BP fixed points}\label{sec:stability}

The next issue to tackle regarding the fixed points of BP is their
stability. The following definition of
conditional belief will be useful
\[
b_{\alpha\beta}(\tau_\alpha|\tau_\beta) \egaldef \frac{b_{\alpha\beta}(\tau_\alpha,\tau_\beta)}{b_\beta(\tau_\beta)}.
\]
For the general case we have the following 
\begin{prop}\label{stability}
  The stability of any fixed point of the BP algorithm is
  determined by the set of beliefs $\{b\}$ of that fixed point: the
  fixed point is stable if, and only if, the matrix defined, for any 
  pair of oriented edges 
  $(\alpha,\beta)\in\E$, $(\alpha',\beta')\in\E$, by the elements
\begin{equation}\label{eq:jacob}
\begin{split}
J_{\alpha\beta}^{\alpha'\beta'}
&=\bigl(b_{\alpha\beta}(1|1) -  b_{\alpha\beta}(1|0)\bigr)
\ind{\alpha'\in \V(\alpha)\setminus\{\beta\},\ \beta'=\alpha}\\
&=\bigl(1-b_{\alpha\beta}(0|1) -  b_{\alpha\beta}(1|0)\bigr)
\ind{\alpha'\in \V(\alpha)\setminus\{\beta\},\ \beta'=\alpha},
\end{split}
\end{equation}
has a spectral radius smaller than $1$.

A sufficient condition for this stability is therefore
\[
 \bigl|b_{\alpha\beta}(1|1) -  b_{\alpha\beta}(1|0)\bigr| <
 \frac{1}{q_\alpha - 1},\text{ for all }\alpha\in\V,\beta\in\V(\alpha).
\]
\end{prop}
\begin{proof}
  Since we are dealing with binary variables, messages are vectors
  with two components, and it is easier to set
\[
\eta_{\alpha\to\beta} \egaldef \frac{m_{\alpha\to\beta}(1)}{m_{\alpha\to\beta}(0)}.
\]

This normalization is equivalent to the one proposed in
Section~\ref{sec:normalization}, according to the change of variables
\[
 \tilde m_{\alpha\to\beta}(0)=\frac{1}{1+\eta_{\alpha\to\beta}}
 \quad\text{and}\quad
 \tilde m_{\alpha\to\beta}(1)=\frac{\eta_{\alpha\to\beta}}{1+\eta_{\alpha\to\beta}},
\]
and the scaled BP algorithm update rule~(\ref{eq:normrule}) can
be rewritten as
\begin{equation}\label{eq:eta}
\eta_{\alpha\to\beta} \gets \frac{b_{\alpha\beta}(0|1) + 
\bigl[\prod_{\gamma\in \V(\alpha)\setminus\{\beta\}}\eta_{\gamma\to\alpha}\bigr]b_{\alpha\beta}(1|1)}
{b_{\alpha\beta}(0|0) + 
\bigl[\prod_{\gamma\in \V(\alpha)\setminus\{\beta\}}\eta_{\gamma\to\alpha}\bigr]b_{\alpha\beta}(1|0)},
\end{equation}
after performing the change of referential of
Proposition~\ref{refchange} with reference point $\{b\}$. We look for
small perturbations around the fixed point
$\eta_{\alpha\to\beta}\equiv1$ for all $(\alpha,\beta)$. 
The Jacobian at the point $\eta=1$ reads:
\[
\frac{\partial \eta_{\alpha\beta}}
     {\partial\eta_{\alpha'\beta'}}\Big|_{\eta=1}
=\bigl(b_{\alpha\beta}(1|1) -  b_{\alpha\beta}(1|0)\bigr)
\ind{\alpha'\in \V(\alpha)\setminus\{\beta\},\ \beta'=\alpha},
\]
which proves~(\ref{eq:jacob}). The rest of the proposition is a
consequence of basic inequalities on the spectral radius of a matrix.
\end{proof}

\begin{figure}[tb]
\parbox[t]{0.33\columnwidth}{\centering%
  \includegraphics[width=0.3\columnwidth]{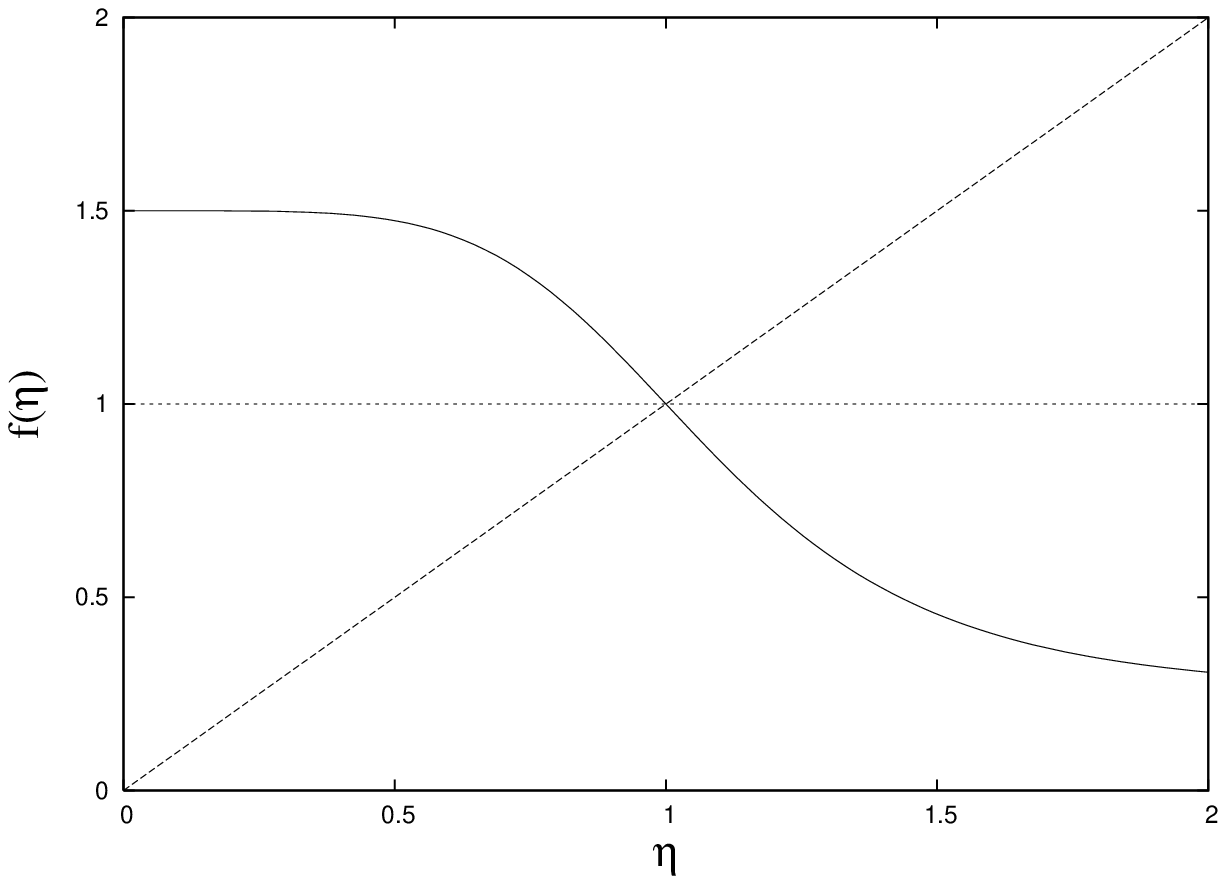}\par
  a) $(q-1)\kappa<-1$%
}\hfill
\parbox[t]{0.33\columnwidth}{\centering%
  \includegraphics[width=0.3\columnwidth]{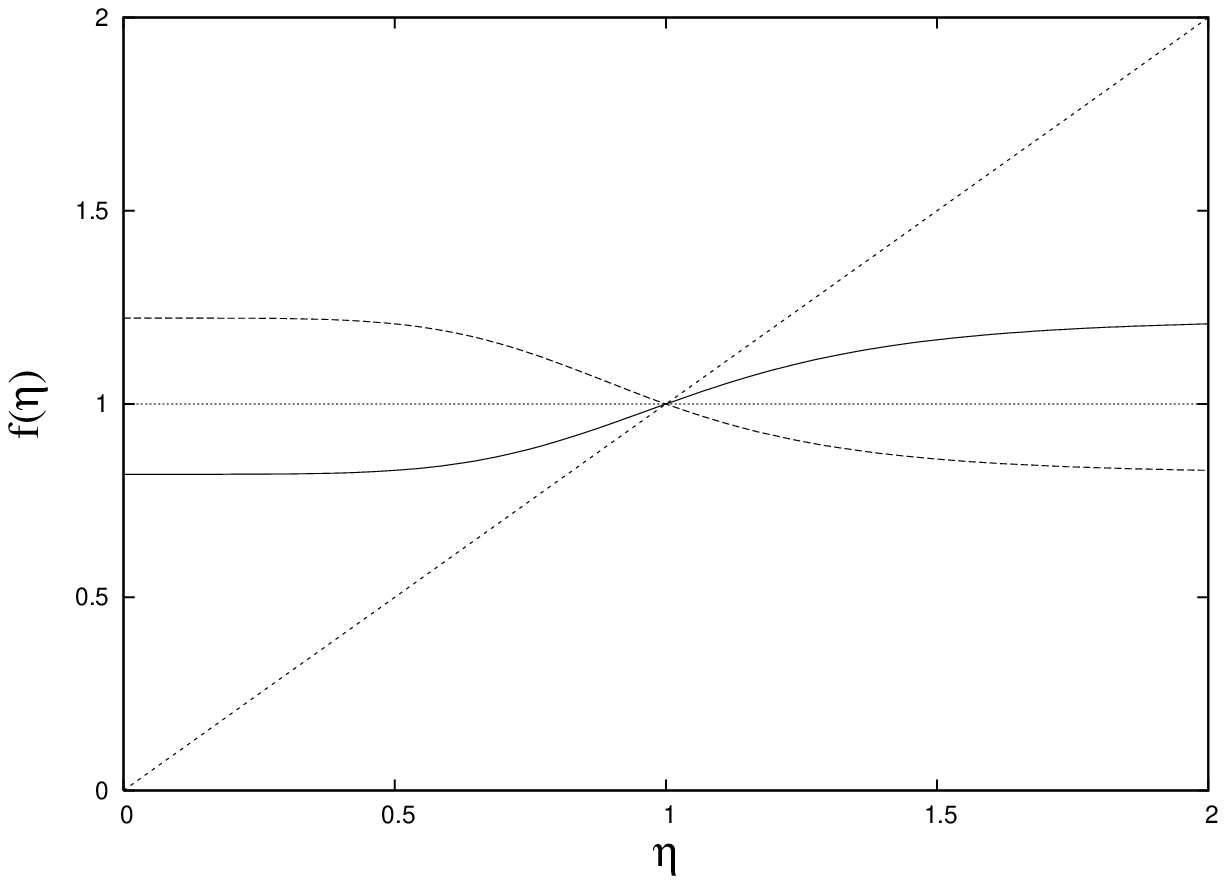}\par
  b) $(q-1)\kappa\in (-1,0)$ and $(q-1)\kappa\in (0,1)$%
}\hfill
\parbox[t]{0.33\columnwidth}{\centering%
  \includegraphics[width=0.3\columnwidth]{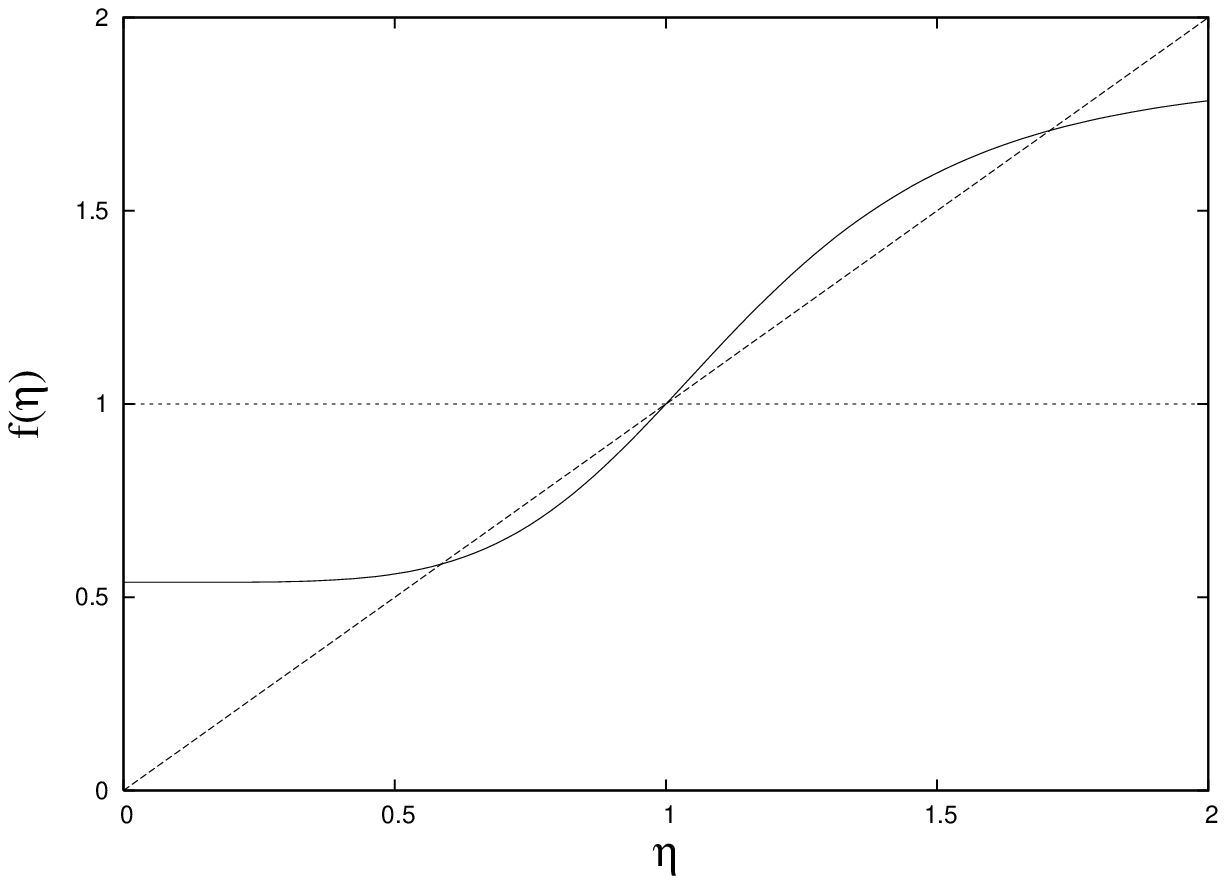}\par
  c) $(q-1)\kappa>1$%
}\hfill
\caption{\label{phase}Different possible graphs of $f(\eta)$ versus
  $\eta$ depending on the value of $\kappa=b(1|1)-b(1|0)$: (a) one
  unstable fixed point, (b) one stable fixed point and (c) one
  unstable and two stable fixed points.}
\end{figure}
\paragraph{Remark}For a totally symmetric graph with connectivity $q$,
(\ref{eq:eta}) reduces to
\[
\eta \gets f(\eta)\egaldef \frac{b(0|1) + \eta^{q-1}b(1|1)}{b(0|0) + \eta^{q-1}b(1|0)},  
\]
and the classification in terms of $b(1|1)-b(1|0)$ is pictured in
Figure~\ref{phase}. Note that $b(1|1)-b(1|0)>0$ (resp.\ $<0$)
corresponds a ferromagnetic (resp.\ anti-ferromagnetic) system.

\bigskip If one considers the dual graph formed by function nodes,
where links relate pairs of function nodes having a variable in
common, then on this graph the Jacobian matrix has the structure of
the incidence matrix $A$ already encountered in the preceding section.
This matrix is not symmetric, but eigenvalues greater than $1$ in
modulus indicate anyway an instability. These are obtained by forming
the new matrix $J(\lambda) \egaldef J - \lambda I$ with $I$ the
identity matrix and finding roots of
\[
\det J(\lambda) = 0.
\]
The expansion of $\det J(\lambda)$ involves permutations which are
compatible with circuits of the dual graph, where each vertex is
visited once. Each permutation is uniquely represented by a product of
permutation cycles (orbits) with disjoint support and is attached to a
sub-graph of the dual graph. Let us call maximal permutation, a
permutation such that the complementary graph of its associated
sub-graph is cycle free.  Adapting results of~\cite{Harary}, $\det
J(\lambda)$ may be expanded according to the following,
\begin{prop}
\begin{equation}\label{Jacobian}
\det J(\lambda) = \sum_{\sigma\in\mathcal{C}}\prod_{\omega_i\subset\sigma} 
(\det \omega_i+(-\lambda)^{|\omega_i|}),
\end{equation}
where the sum runs over all possible maximal  permutations $\sigma$, 
each one being expressed as a product of
$n\ge 1$ circular permutations (cycles) $\omega_i,i=1\ldots n$, of
size $|\omega_i|$,
with determinant  given by 
\[
\det \omega = -(-1)^{|\omega|}\prod_{(\alpha,\beta)\in\omega}
\bigl(b_{\alpha\beta}(1|1) - b_{\alpha\beta}(1|0)\bigr).
\]
\end{prop}
On a tree, as expected, zero is the only eigenvalue, in fact $J$ is a
nilpotent matrix of index the size of the longest directed path in the
graph. If there is only one cycle $\omega$, (\ref{Jacobian}) reduces
to
\[
\det J(\lambda) = \lambda^{N-|\omega|}(\det \omega+(-\lambda)^{|\omega|}),
\]
which yields the eigenvalues
\[
\lambda_k = \biggl(\prod_{(\alpha,\beta)\in\omega}
\bigl(b_{\alpha\beta}(1|0) -  b_{\alpha\beta}(1|1)\bigr)
\biggr)^{\frac{1}{\omega}}e^{(2k+1)i\pi/|\omega|},
\]
with modulus obviously smaller than one.
As a consequence, the following proposition holds.
\begin{prop}
BP fixed points for a graph containing at most one oriented loop are stable.
\end{prop}
This has been remarked by different means in~\cite{Heskes3}.
Unstable modes correspond to eigenvalues larger than $1$, and might
reveal vertices or cycles mostly responsible for the instabilities. An interesting case
occurs when cycles of the dual graph have disjoint supports, because then
only one maximal permutation $\sigma$ exists and expansion~(\ref{Jacobian}) reduces to one term,
\[
\det J(\lambda) = \prod_{\omega_i\subset\sigma} 
(\det \omega_i+(-\lambda)^{|\omega_i|}).
\]
As a result, since the modulus of the Jacobian coefficients are always
smaller than $1$, to each cycle is associated an eigenvalue smaller
than $1$ and the state is stable.

On a
graph which is locally a tree (Bethe lattice), a mean-field equation
can be used to evaluate the stability of a given fixed point. The idea is
to consider the iterated Jacobian matrix in a statistical manner, by
looking at the distribution $P^{(n)}(v)$ of components $v$ of an
iterated vector starting from a non-degenerate initial condition,
\[
V^{(n)} = J^n V^{(0)}.
\]
The mean-field stability equation then simply reads (after assuming
the usual independence property of parent messages)
\begin{equation}\label{eq:mean-field}
P^{n+1}(v) = \sum_{c>1}Q(c)\sum_{\{v_i\},\{\kappa_i\}}\prod_{i=1}^{c-1}
P^{n}(v_i)R(\kappa_i)\delta(v-\sum_{i=1}^{c-1} v_i\kappa_i),
\end{equation}
with $Q$ the connectivity distribution in the dual graph and $R$ the Jacobian
coefficient distribution (see Figure~\ref{histo} for example). The instability is
therefore fully characterized by the statistical properties of the
considered BP fixed point and by the statistical properties of the
graph (connectivity), which sometimes can be an adjustable parameter.

\section{Toy Model simulations}\label{simulation}

\subsection{From theory to practice}

We illustrate these ideas on a simulated traffic system which has the advantage
to yield exact empirical data correlations. For real data, problems may
arise because of noise in the historical information used to build the 
model. This additional difficulty will be treated in a separate work. 

The model consists of a queueing network system.  Each queue represents
a link of the traffic network (a single-way lane) and has a finite
capacity; to each link we attach a variable $\rho\in[0,1]$,
the car density, which is represented by a color code in the picture
(Figure~\ref{fig:photo} on page~\pageref{fig:photo}).

As already stated in Section~\ref{traffdes}, the physical traffic
network is replicated, to form a space time graph, in which each
vertex $\alpha = (\ell,t)$ corresponds to a link $\ell$ at a given time $t$
of the traffic graph. To any space-time vertex $\alpha$, we associate
a binary congestion variable $\tau_\alpha\in\{0,1\}$. 

The statistical physics description amounts to relating the
probability of saturation $P(\tau_\alpha = 1)$ to the density
$\rho_\alpha$. For the sake of simplicity, we consider a linear relation
and build our historical $\hat p$ according to the
rules
\begin{align*}
\hat p_\alpha(1) &= \mu_{\infty}(\rho_\ell),\\[0.2cm]
\hat p_\alpha(0) &= \mu_{\infty}(1-\rho_\ell),\\[0.2cm]
\hat p_{\alpha\beta}(1,1) &= \mu_{\infty}(\rho_\ell\rho_{\ell'}),\\[0.2cm]
\hat p_{\alpha\beta}(1,0) &= \mu_{\infty}(\rho_\ell(1-\rho_{\ell'})),\\[0.2cm]
&\ldots
\end{align*}
where $\mu_{\infty}$ is simply a frequency estimator. Note that, to
follow some realistic statistical constraints, we use here only 
data aggregated in time.
More realistic data collection and modeling would work the same way.
 
\begin{figure}
\resizebox*{0.5\textwidth}{!}{\input{meta_graph.pstex_t}}
\resizebox*{0.4\textwidth}{!}{\input{incidence.pstex_t}}
\caption{Structure of the factor graph (left), 3 time-layers portions are represented, 
black circles correspond to crossroads and blue squares to factor-vertices.  
Corresponding graph for the Jacobian matrix (right).}\label{meta}
\end{figure}
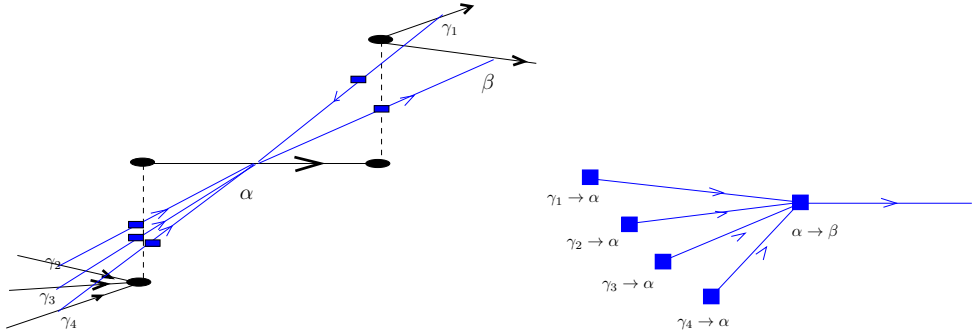

The structure of the factor-graph on which we propagate the information 
is depicted in Figure~\ref{meta}.

\begin{figure}[p]
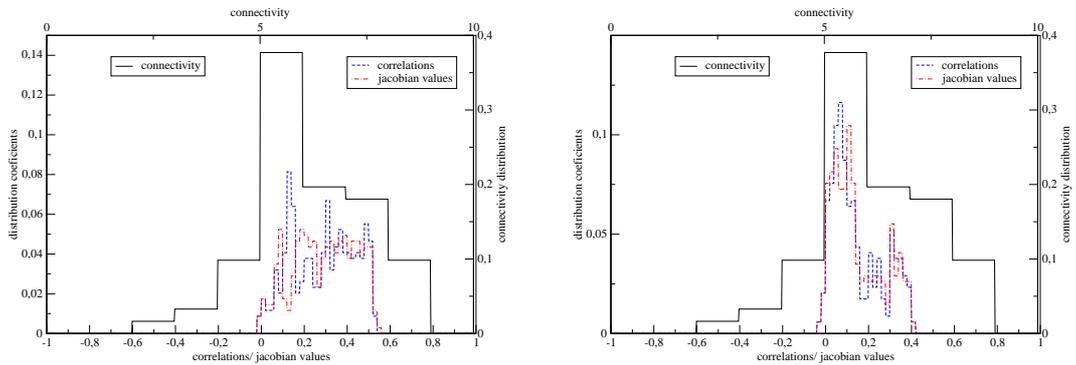

\centering
\includegraphics*[width=0.47\textwidth]{histo_20}\hfill
\includegraphics*[width=0.47\textwidth]{histo_40}
\caption{\label{histo}Connectivity (black), correlation coefficients (red) and 
  Jacobian coefficients (blue) histograms for low (top) and high
  (bottom) noise level.}
\end{figure}

Some fine tuning is required to let the algorithm work correctly. 
First, from Proposition~\ref{stability}, we
know that the stability of the reference point $\hat p$ encoded in
(\ref{bethepsi})--(\ref{bethephi}) is not guaranteed; this may be
evaluated on the basis of distributions depicted in
Figure~\ref{histo}, using equation (\ref{eq:mean-field}). In absence
of negative correlations, it is likely that our system is either a
paramagnetic-like (in the Ising-model terminology) system, with small
fluctuations around average values, or a ferromagnetic-like system, in
the sense that positive correlation drive the system to a state
where links are in a similar state, i.e.\ mostly fluid (low state) 
or congested (high state).
This scenario corresponds to the regimes pictured on Figure~\ref{phase}
where case (c) is the usual ferromagnetic phase transition
in the Ising model. It is also a well-known fact that this transition
is driven by the temperature. To introduce the equivalent of a
temperature in our equations, since its effect is essentially to
reduce correlations, let us consider modified pairwise marginal laws
\[
\tilde p_{\alpha\beta}(\epsilon) = \epsilon \hat
p_{\alpha\beta}+(1-\epsilon)\hat p_{\alpha}\hat p_{\beta}.
\]
The high temperature regime corresponds here to $\epsilon\to 0$ and
the vanishing of the correlations. The Jacobian
coefficients $\kappa_{\alpha\beta} \egaldef b_{\alpha\beta}(1|1) -
b_{\alpha\beta}(1|0)$ are modified according to
\[
\hat \kappa_{\alpha\beta}(\epsilon) = \epsilon \kappa_{\alpha\beta},
\]
which means that eigenvalues are rescaled by a factor $\epsilon$. For
our purpose, this provides us with an adjustable mean-field parameter,
to correct some artificial amplification of correlations caused by
closed loops in the graph. We expect that there exists a critical
value of $\epsilon_c$ corresponding to the ferromagnetic phase
transition point (high temperature means here small $\epsilon$).  In
addition, since for small $\epsilon$ we recover in one sweep the bare
mean results, this parameter can be used for a simulated annealing
procedure, by letting it converge from zero to the desired value
during the BP iterations.

The second adjustment concerns the encoding of real-time information.
The probe vehicle is assumed to send an information for some
space-time vertex $\alpha^*$, typically in the form of an
instantaneous velocity, from which is estimated the probability
$p_{\alpha^*}$ of saturation. Instead of projecting this information
on one of the two states ($\tau_{\alpha^*}=0$ or $\tau_{\alpha^*}=1$),
which turns out in practice to be too coarse, we use a procedure which
amounts to bias the messages sent by $\alpha^*$ in proportion to the
observed belief $p_{\alpha^*}$. In the statistical physics language,
this amounts to impose an external local field on the observed
variable.

The last issue concerns the situation where the system is below the
transition point, in which case we have two separate states, and it is
always possible that BP converges towards the wrong fixed-point.
In this simple ferromagnetic situation, it is in fact easy to enforce the
convergence of the algorithm to a specified  fixed point by applying a slowly
decaying  external field,
enforcing either the fluid or the congested state. As a result of this
procedure, we obtain two sets of beliefs $\{b^0\}$ and $\{b^1\}$, with
corresponding free energies $F^0$ and $F^1$, from which we build
the superposition belief,
\[
b_\alpha = \frac{e^{-F^0}b_\alpha^0+e^{-F^1}b_\alpha^1}{e^{-F^0}+e^{-F^1}}.
\]  
which in practice, for sufficiently large systems, because $F$ is extensive, turns out to 
be the set of beliefs corresponding to the lowest free energy.

In the following we refer the sets of belief $\{b_\alpha^0\}$, $\{b_\alpha^1\}$ and
$\{b_\alpha\}$ respectively to the \emph{low}, the \emph{high} 
and the \emph{combined inference state}. Accordingly the set $\{\hat p_\alpha\}$ is 
referred to as the \emph{historical state}. In addition, the combination of observations
with  historical data (by replacing the historical value with the last observation 
in the window time) yields the \emph{actual state}. 
To estimate the quality of the traffic restoration we use the following estimator:
\[
\text{reconstruction rate} \egaldef \frac{1}{|\V|}\sum_{\alpha\in\V}\ind{|b_\alpha-\rho_\alpha|<0.2},
\]
which computes the fraction of space-time
nodes $\alpha$ for which the belief $b_\alpha$ does not differ by more than
an arbitrary threshold of $0.2$ from $\rho_\alpha$. 

\subsection{Numerical results}

\begin{figure}[p]
\centering
\includegraphics[width=0.7\columnwidth]{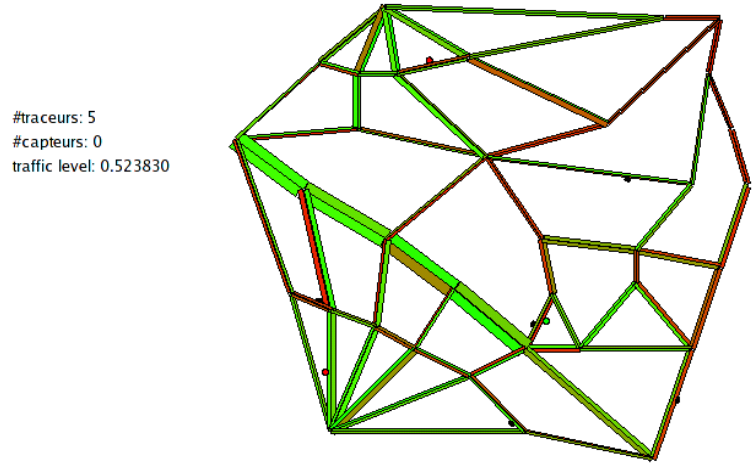}
\caption{Traffic network as produced by the simulator.
  The continuous color code represents the traffic index from 0 (full
  green) to 1 (full red).}\label{fig:photo}
\end{figure}

\begin{table}[p]
\centering
\begin{tabular}{|c|c|c|c||c|c|}
\hline
 nodes & links & time steps & graph size & $\epsilon_{c_1} $(oscillating) 
& $\epsilon_{c_2}$ (noisy)\\[0.2cm]
\hline 
 35 & 122 & 43 & 5246 & 0.67  & 1.29\\
\hline  
\end{tabular}
\caption{Toy model characteristics}\label{caract}
\end{table}

\begin{figure}
\centering
\includegraphics*[width=0.7\textwidth]{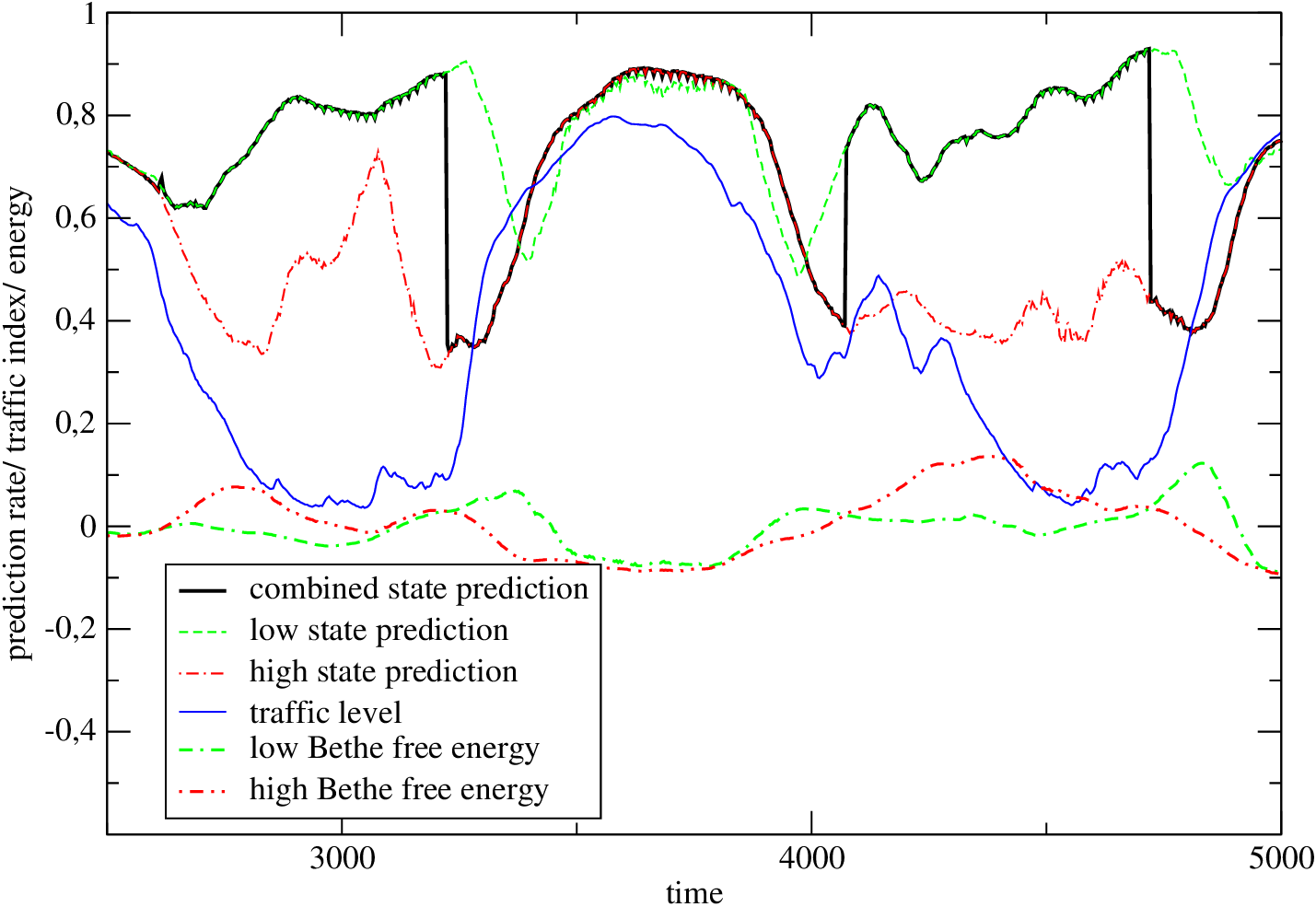}
\caption{\label{t_serie} Reconstruction rates for the various possible 
inference states as a function of time with corresponding free energies, with $10$
probe vehicles and $\epsilon=0.75$.}

\bigskip
\includegraphics*[width=0.7\textwidth]{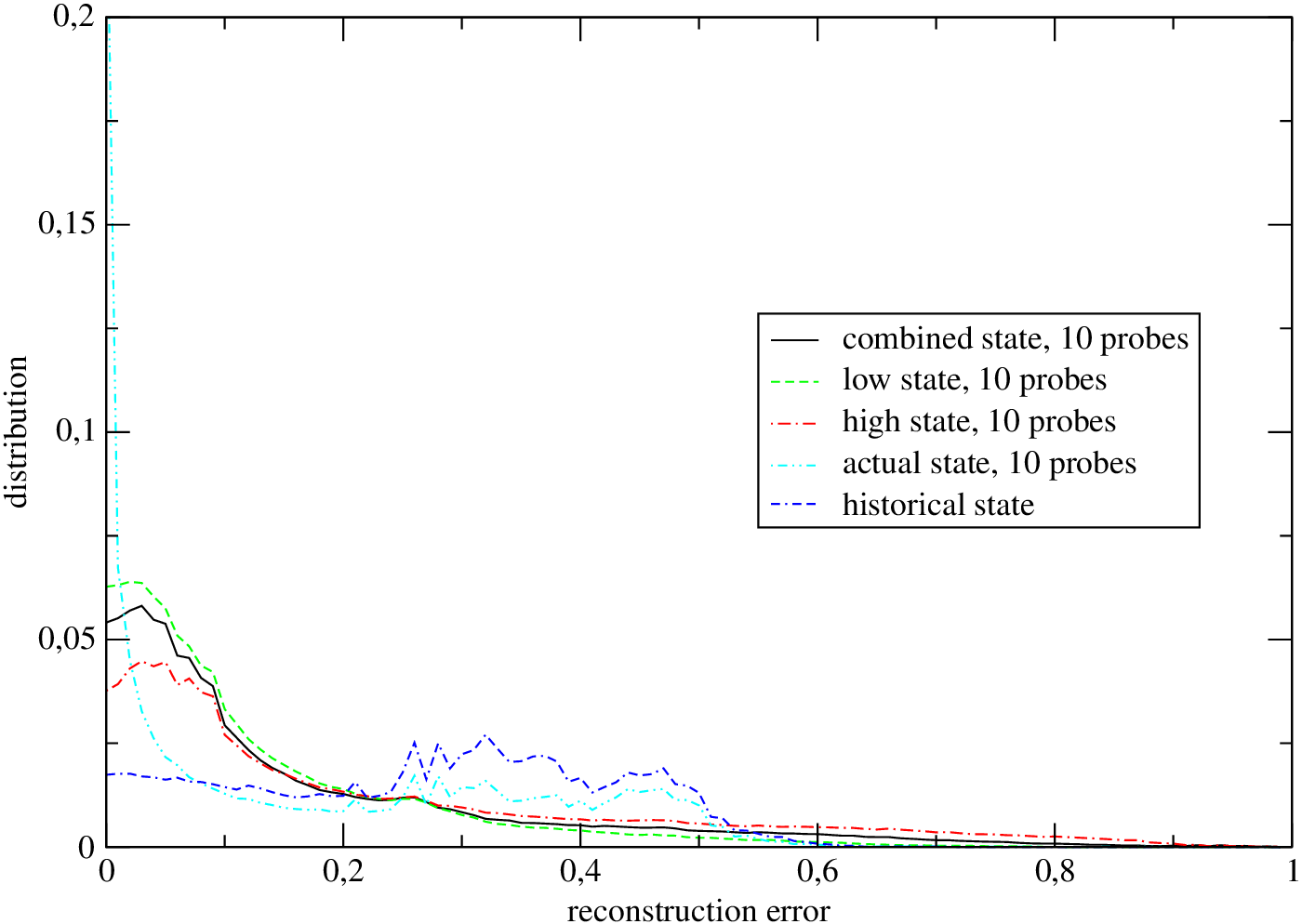}
\caption{\label{perf_distrib} Distribution of reconstruction errors for the various possible 
inference states for $\epsilon=0.75$.}
\end{figure}

We have tested the algorithm on the toy traffic network shown on the
program's screen-shot of Figure~\ref{fig:photo}. The characteristics of
this network are summarized in Table~\ref{caract}. Two types of
traffic conditions have been used, that both correspond to periodic
oscillation superimposed with noise (see blue curve of
Figure~\ref{t_serie}); they simply differ by the level of the noise.

The two values of $\epsilon_c$ in Table~\ref{caract} that have been
computed for the two different traffic regimes
using~(\ref{eq:mean-field}) are close to the observed values, which
indicates that the space-time graph on which BP is run is close to
the conditions of a dilute graph (Bethe lattice).

The simulation run of Figure~\ref{t_serie} compares the policies of
using only the low state (green), the high state (red) or the combined
state w.r.t.\ the free energy in the low-noise case. Abrupt changes of the 
combined state prediction correspond to the crossing of the Bethe free energies.
In the transition regimes, which correspond to out-of-equilibrium situations,
the free energy criteria sometimes select the wrong state. The
reason for this is that the present design of our algorithm encodes
only statistical information at equilibrium.  Time correlations should be
incorporated in some way, to encode transition rates between the
macro-states (here the low and high traffic density).

\begin{figure}
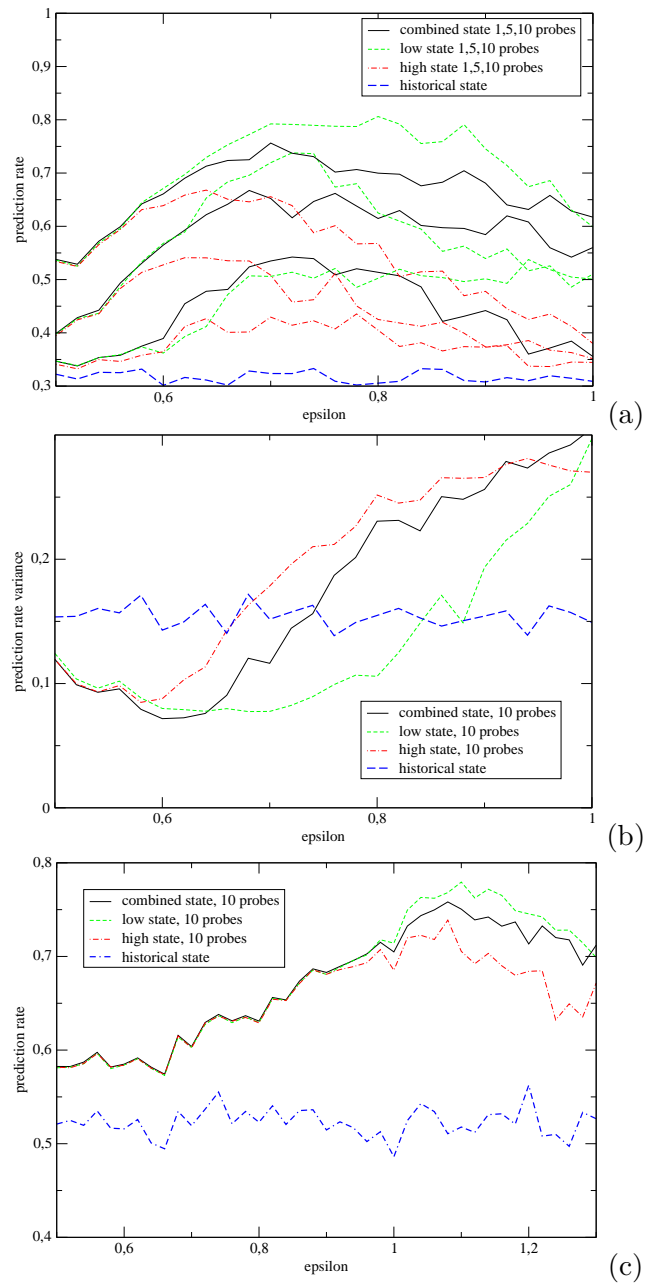

\centering
\includegraphics*[width=0.55\textwidth]{perf_beta_20}
(a)
\includegraphics*[width=0.55\textwidth]{var_beta_10_20}
(b)
\includegraphics*[width=0.55\textwidth]{perf_beta_40}
(c)
\caption{\label{perf_epsilon} (a) Reconstruction rates obtained with $0$, $1$,
$5$ and $10$ probe vehicles and various possible inference states; 
 (b) variance of the reconstruction rate obtained with $10$ vehicles
also for the various possible states; (c) reconstruction rates obtained for 
the noisy network, again with $10$ probes.}
\end{figure}
\begin{figure}
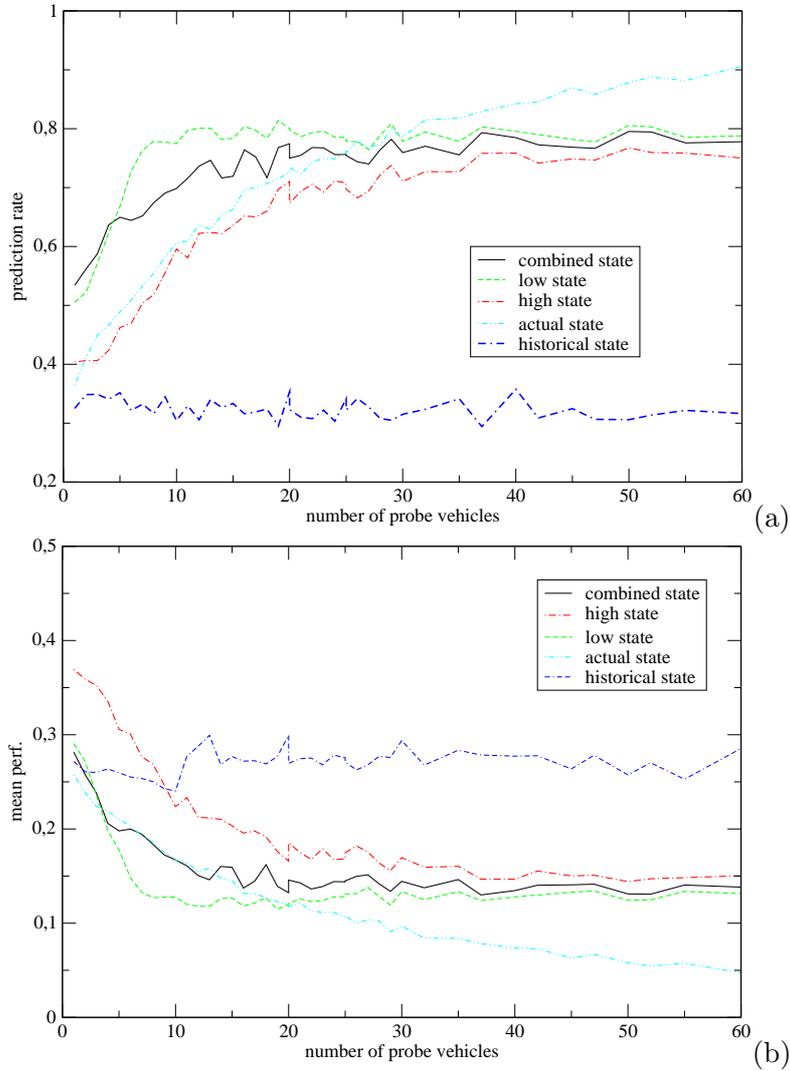

\centering
\includegraphics*[width=0.7\textwidth]{prediction_nt_0.78_20.eps}(a)
\centering
\includegraphics*[width=0.7\textwidth]{perf_nt_0.78_20.eps}(b)
\caption{\label{perf_nt} (a) reconstruction rates with various possible 
inference states when the number of probe vehicles is varied; (b)
corresponding average prediction error.}
\end{figure}

Distributions of performance errors shown in
Figure~\ref{perf_distrib} are based on a simulation run of $10000$
traffic time units where a belief propagation is run every 3 units of
time for both low and high inference states, to reconstruct the
traffic.  Varying the parameters (either $\epsilon$ or the number of
probe vehicles) and integrating the distributions up to $0.2$ yields
curves of Figures~\ref{perf_epsilon} and~\ref{perf_nt}.
They indicate that the optimal value of $\epsilon$ for traffic
prediction is slightly above the critical value for the traffic oscillating network
and below the critical value for the noisy network, as expected. The fact that the
prediction rate saturates at $0.8$ when the number of probe vehicles
is increased in Figure~\ref{perf_nt} is again due to the traffic
transition regimes.



\section{Conclusion and perspectives}\label{sec:conclusion}

We have presented a novel methodology for reconstruction and
prediction of traffic using the Belief Propagation algorithm on
Floating Car Data. We have shown how the underlying Ising model can be
determined in a straightforward manner and that it is unique up to
some change of variables. In addition, the effect of message
normalization and the stability properties can be asserted from the
original data. The unfortunate fact that the BP fixed point
corresponding to the historical data may be unstable can be
circumvented by rescaling of the correlations. The algorithm has been
implemented and illustrated using a toy traffic model.

Several generalizations are considered for future work:
\begin{itemize}
\item firstly, the binary description corresponding to the underlying
  Ising model is arbitrary. Traffic patterns could be represented in
  terms of $p$ different inference states.  A Potts model with
  $p$-states variables would leave the belief propagation algorithm
  and its stability properties structurally unchanged. Actually this
  number $p$ should be subject to an optimization procedure.  
\item secondly, our way of encoding traffic network information might
  need to be augmented to cope with real world situations. This would
  simply amount to redefine the factor-graph used to propagate this
  information. In particular it is likely that a great deal of
  information is contained in the correlations of local congestion
  with aggregate traffic indexes, corresponding to sub-regions of the
  traffic network. Taking these correlations into account would result
  in the introduction of specific variables and function nodes
  associated to these aggregate traffic indexes. These aggregate
  variables would naturally lead to a hierarchical representation of
  the factor graph, which is necessary for inferring the traffic on
  large scale network. Additionally, time dependent correlations which
  are needed for the description of traffic, which by essence is an
  out of equilibrium phenomenon, could be conveniently encoded in
  these traffic index variables.
\end{itemize}

Ultimately, for the elaboration of a powerful prediction system, 
the structure of the information content of a traffic-road network 
has to be elucidated through a specific statistical analysis. The use 
of probe vehicles, based on modern communications devices, combined with a
belief propagation approach, is  in this respect a very promising approach.
 
\bibliography{refer}
\bibliographystyle{amsplain}

\end{document}

%% file: meta_graph.pstex_t
\begin{picture}(0,0)%
\includegraphics{meta_graph.pstex}%
\end{picture}%
\setlength{\unitlength}{3947sp}%
\begingroup\makeatletter\ifx\SetFigFont\undefined%
\gdef\SetFigFont#1#2#3#4#5{%
  \reset@font\fontsize{#1}{#2pt}%
  \fontfamily{#3}\fontseries{#4}\fontshape{#5}%
  \selectfont}%
\fi\endgroup%
\begin{picture}(5840,3640)(1009,-4308)
\put(3583,-2857){\makebox(0,0)[lb]{\smash{{\SetFigFont{14}{16.8}{\rmdefault}{\mddefault}{\updefault}$\alpha$}}}}
\put(6237,-1637){\makebox(0,0)[lb]{\smash{{\SetFigFont{14}{16.8}{\rmdefault}{\mddefault}{\updefault}$\beta$}}}}
\put(5800,-1006){\makebox(0,0)[lb]{\smash{{\SetFigFont{12}{14.4}{\rmdefault}{\mddefault}{\updefault}$\gamma_1$}}}}
\put(1448,-3599){\makebox(0,0)[lb]{\smash{{\SetFigFont{12}{14.4}{\rmdefault}{\mddefault}{\updefault}$\gamma_2$}}}}
\put(1387,-3986){\makebox(0,0)[lb]{\smash{{\SetFigFont{12}{14.4}{\rmdefault}{\mddefault}{\updefault}$\gamma_3$}}}}
\put(1621,-4250){\makebox(0,0)[lb]{\smash{{\SetFigFont{12}{14.4}{\rmdefault}{\mddefault}{\updefault}$\gamma_4$}}}}
\end{picture}%

%% file: incidence.pstex_t
\begin{picture}(0,0)%
\includegraphics{incidence.pstex}%
\end{picture}%
\setlength{\unitlength}{3947sp}%
\begingroup\makeatletter\ifx\SetFigFont\undefined%
\gdef\SetFigFont#1#2#3#4#5{%
  \reset@font\fontsize{#1}{#2pt}%
  \fontfamily{#3}\fontseries{#4}\fontshape{#5}%
  \selectfont}%
\fi\endgroup%
\begin{picture}(4287,1642)(2131,-4436)
\put(2683,-4008){\makebox(0,0)[lb]{\smash{{\SetFigFont{10}{12.0}{\rmdefault}{\mddefault}{\updefault}$\gamma_3\to\alpha$}}}}
\put(2329,-3580){\makebox(0,0)[lb]{\smash{{\SetFigFont{10}{12.0}{\rmdefault}{\mddefault}{\updefault}$\gamma_2\to\alpha$}}}}
\put(2131,-3135){\makebox(0,0)[lb]{\smash{{\SetFigFont{10}{12.0}{\rmdefault}{\mddefault}{\updefault}$\gamma_1\to\alpha$}}}}
\put(4594,-3473){\makebox(0,0)[lb]{\smash{{\SetFigFont{10}{12.0}{\rmdefault}{\mddefault}{\updefault}$\alpha\to\beta$}}}}
\put(3441,-4387){\makebox(0,0)[lb]{\smash{{\SetFigFont{10}{12.0}{\rmdefault}{\mddefault}{\updefault}$\gamma_4\to\alpha$}}}}
\end{picture}%